\documentclass[aps,reprint,twocolumn,superscriptaddress,citeautoscript,showpacs,amsmath,amssymb,floatfix,pre]{revtex4-2}

\usepackage{graphicx}
\usepackage{xcolor}
\usepackage{epstopdf}
\usepackage{amsmath,amsthm,amssymb}
\usepackage{latexsym}
\usepackage{bm}
\usepackage{dcolumn}
\usepackage{braket}
\usepackage[normalem]{ulem}
\usepackage{times}
\usepackage{hyperref}
\newcommand{\revblue}[1]{\textcolor{black}{#1}}
\makeatletter
\DeclareRobustCommand{\rev}{\@ifnextchar1{\rev@one}{\revblue}}
\def\rev@one1#1{\textcolor{black}{#1}}
\makeatother
\begin{document}

\title{Evolutionary chemical learning in dimerization networks}

\author{Alexei V. Tkachenko}
\email{oleksiyt@bnl.gov}
\affiliation{Center for Functional Nanomaterials, Brookhaven National Laboratory, Upton, NY 11973, USA}
\author{Bortolo Matteo Mognetti}
\affiliation{Interdisciplinary Center for Nonlinear Phenomena and Complex Systems, Universit\'e Libre de Bruxelles, B-1050 Brussels, Belgium}
\author{Sergei Maslov}
\email{maslov@illinois.edu}
\affiliation{Department of Bioengineering, University of Illinois Urbana-Champaign, Urbana, IL 61801, USA}
\affiliation{Carl R. Woese Institute for Genomic Biology, University of Illinois Urbana-Champaign, Urbana, IL 61801, USA}
\affiliation{Department of Physics, University of Illinois Urbana-Champaign, Urbana, IL 61801, USA}

\begin{abstract}
We present a  framework for chemical learning based on Competitive Dimerization Networks (CDNs)— systems in which multiple molecular species, e.g., proteins, DNA oligomers, or RNA oligomers, reversibly bind to form dimers. We \rev1{show numerically that these networks can, in principle, be trained \emph{in vitro}} through directed evolution, enabling the implementation of complex learning tasks such as multiclass classification without digital hardware or \rev1{prior knowledge of all microscopic association constants}. Each molecular species functions analogously to a neuron, with binding affinities acting as tunable synaptic weights. A training protocol involving mutation, selection, and amplification of DNA-based components allows CDNs to robustly discriminate among noisy input patterns. The resulting classifiers exhibit strong output contrast and high mutual information between input and output, especially when guided by a contrast-enhancing loss function. Comparative analysis with in silico gradient descent training reveals closely correlated performance. These results establish CDNs as a promising platform for analog physical computation, bridging synthetic biology and machine learning, and advancing the development of adaptive, energy-efficient molecular computing systems.
\end{abstract}

\maketitle

\section{Introduction}

Deep learning has transformed science, technology, and everyday life by enabling artificial intelligence systems to extract patterns and make complex decisions from large datasets. Traditional approaches rely on digital computation and optimization techniques, such as backpropagation and gradient descent, to train models like deep neural networks (DNNs) \cite{krizhevsky2012imagenet,Goodfellow2016}. However, the high energy cost and limited scalability of digital architectures have motivated the search for \emph{analog neuromorphic systems}, in which physical processes themselves perform computation. Optical, electronic, and mechanical implementations of artificial neural networks have shown that information processing can emerge directly from intrinsic system dynamics \cite{optical,photonics}. 

A further evolution of this idea is the concept of \emph{physical learning}, where the system not only performs computation but can also \emph{learn} desired functions through self-adjustment of its physical parameters, rather than through explicit programming \cite{memristors,Liu2021PRX,phys_intelligence,Arvind2015,Wright2022,Science2023,Arvind2023}. Examples include adaptive memristive networks, photonic learning architectures, and mechanical assemblies capable of gradient-free optimization. These developments raise a central question: can matter itself be organized to learn in the physical sense?

Biochemical networks in living cells provide a compelling natural example of physical learning. Processes such as cellular signaling, gene regulation, and metabolic control display remarkable adaptability, continuously optimizing responses to environmental cues \cite{bray1995protein,bhalla1999emergent,kramer2022multimodal}. This intrinsic ability to process and integrate information suggests a general paradigm of \emph{Chemical Learning} (CL), where molecular interactions implement computations analogous to those in artificial neural networks \cite{poole2017dna-860,Suri2024}. 

Here we introduce a CL framework based on \emph{Competitive Dimerization Networks} (CDNs), systems composed of multiple molecular species capable of reversible pairwise binding, where each molecule can form dimers with multiple partners \cite{Maslov2007,Maslov2007njp,Nikitin2023,parresgold2025contextual-aac,brannetti2025covalent-c8c}. Such networks are ubiquitous in biology, from receptor–ligand interactions to transcription-factor binding, and can also be engineered using synthetic molecules. CDNs based on partially complementary DNA oligomers were recently realized experimentally \cite{Nikitin2023}, providing a platform where the association constants are sequence-dependent and can be tuned using well-established nearest-neighbor thermodynamic models \cite{santalucia1998unified-7e0}. 

Classical DNA computing has already demonstrated that nucleic-acid reaction networks can perform logic operations, digital computation, and even emulate arbitrary chemical kinetics \cite{Adleman1994,Seelig2006,Soloveichik2010,Srinivas2017}. In most of these schemes, computation relies on perfect complementarity and strand displacement. In contrast, the partial complementarity of Ref.~\cite{Nikitin2023} enables programmable many-to-many binding, precisely matching the rules of the CDN model. Further advances in covalent and dynamic DNA architectures have expanded the design space, enabling chemical circuits that translate multichannel inputs into programmable dimer outputs \cite{brannetti2025covalent-c8c}. In parallel, protein-based competitive dimerization networks have been explored for contextual and many-to-many computation in living cells \cite{Klumpe2023,parresgold2025contextual-aac}. Together, these developments identify both DNA- and protein-based CDNs as versatile physical substrates for adaptive molecular computation. 

Building on this foundation, we focus on the learning aspect, adapting molecular interactions from examples rather than encoding fixed logic. We introduce a framework for \emph{evolutionary chemical learning} based on CDNs, in which each molecule acts analogously to a neuron, and the binding affinities play the role of tunable synaptic weights. Unlike traditional biochemical circuits with pre-defined architectures, CDNs can be trained either \emph{in silico} through computational optimization or \emph{in vitro} via directed evolution. The latter constitutes a form of physical learning achieved through cycles of mutation, selection, and amplification, allowing molecular systems to acquire functionality \rev1{without a complete prior characterization of the microscopic binding parameters}. 

\rev1{Evolutionary optimization of neural networks and genetic algorithms have previously been employed as numerical tools \cite{Stanley2019Neuroevolution,parresgold2025contextual-aac}. Our proposal takes a further step toward physical learning by formulating an \emph{in vitro} directed-evolution protocol for training the molecular network itself. This approach builds on the strong and well-established foundations of directed evolution in biological research \cite{Wang2021DirectedEvolution}.  }

\rev{A key conceptual distinction of our approach is that CDNs are treated not merely as engineered molecular circuits with prescribed functions, but as \emph{learning systems} whose functionality is acquired from data. While prior studies primarily focus on designing CDNs to implement specific input--output mappings, here we formulate CDNs within a supervised learning framework, enabling adaptation through loss-driven optimization. This shift from engineered computation to learned functionality places CDNs within the emerging paradigm of physical learning.}

\rev{Although DNA-based implementations provide a concrete and experimentally grounded realization, our theoretical framework is platform-agnostic. The concept of chemical learning introduced here applies equally to protein-based networks or other molecular systems capable of reversible binding, substantially broadening the scope and relevance of the approach.}

Beyond establishing chemical learning as a bridge between digital, physical, and biological computation, CDN-based architectures may ultimately be integrated into engineered nanosystems, serving as adaptive elements in functional or responsive self-assembly, biosensing, or smart drug-delivery platforms.

\section{Model and Protocols}
\subsection{Competitive Dimerization Networks}

\begin{figure}
    \centering
    \includegraphics[width=\linewidth]{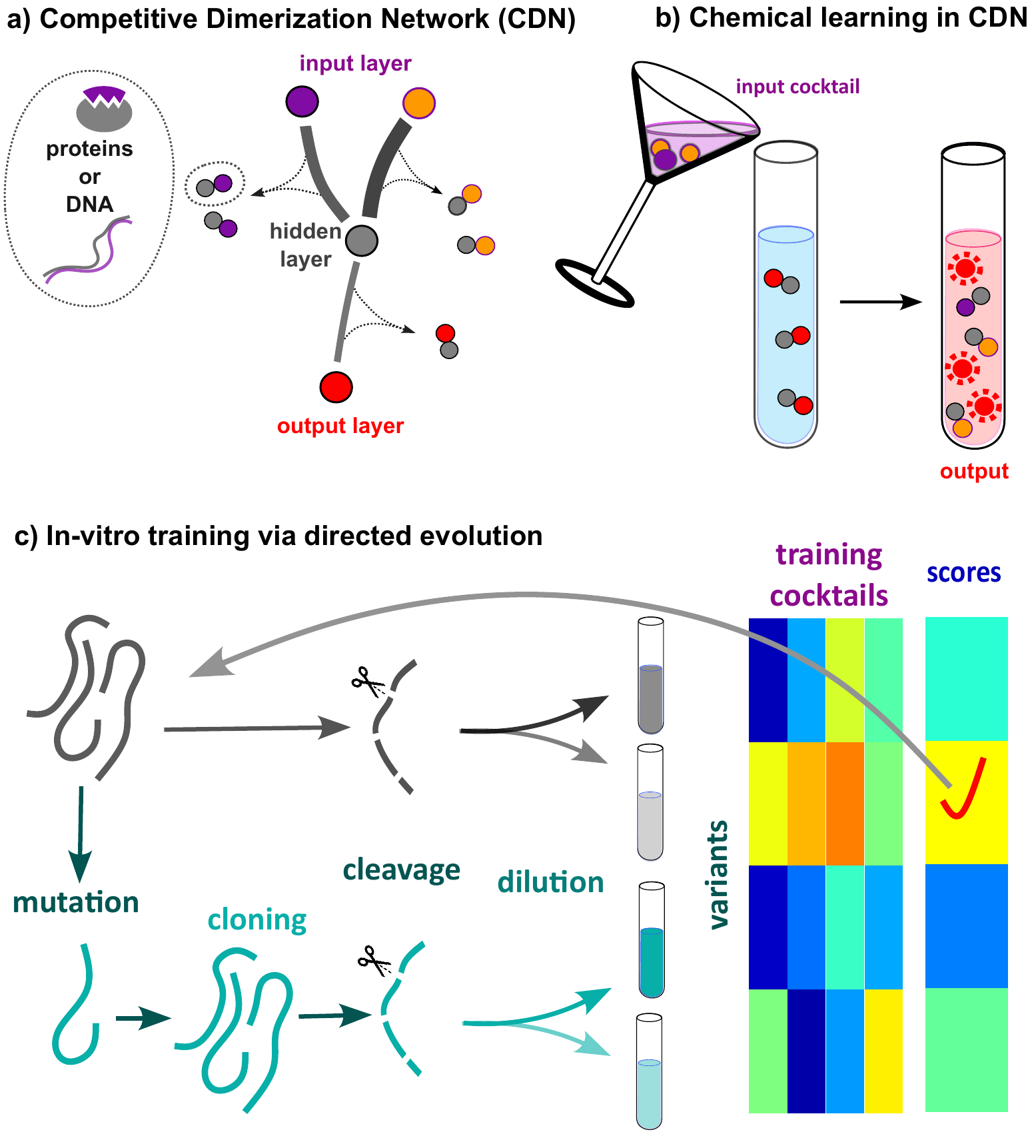}   
    \caption{Overview of chemical learning in competitive dimerization networks. (a) Schematic of a network composed of molecular species that reversibly form pairwise dimers, with designated input, hidden, and output layers. (b) Illustration of chemical computation: input cocktails modulate mass action binding equilibria to generate output fugacities. (c) In vitro training protocol using directed evolution, involving mutation, bond cleavage, dilution, and selection of DNA sequences and their concentrations to optimize classification performance.}
    \label{fig:Fig1}
\end{figure}

In our model, a \textbf{competitive dimerization network (CDN)} consists of \(N\) distinct molecular species (e.g., DNAs or proteins) that can form pairwise dimers \rev1{(two-molecule complexes; in a DNA realization, partially base-paired duplexes)} (see Fig. \ref{fig:Fig1}a). The formation of complexes larger than dimers is assumed to be negligible. We denote the overall concentrations of these molecules by \(\mathbf{c} = (c_1, c_2, \dots, c_N)\) and their fugacities––the fractions of total concentration that remain unbound––by \(\mathbf{f} = (f_1, f_2, \dots, f_N)\).

In chemical mass-action equilibrium free concentrations (activities) of the molecules, $x_i=f_ic_i$ must satisfy the following set of equations:
\begin{equation}
c_i = x_i + \sum_{j} K_{ij} x_ix_j
\label{eq:X}
\end {equation}
Here, the symmetric association constant \(K_{ij} = K_{ji}\) characterizes the strength of dimer formation between molecules of types \(i\) and \(j\). \rev1{The first term on the right-hand side of Eq.~\ref{eq:X} is the free concentration of species $i$, whereas $K_{ij}x_ix_j$ is the concentration of species $i$ bound in an $i$--$j$ dimer; the sum therefore accounts for all dimers containing species $i$. In this work we set $K_{ii}=0$, excluding self-interactions.} The above equilibrium conditions  yield the following equations for fugacities $f_i$: 
\begin{equation}
f_i = \frac{1}{1 + \sum_{j} K_{ij} c_j f_j} ,
\label{eq:f}
\end{equation}
Despite its relative simplicity, we demonstrate below that the CDN is capable of performing learning tasks analogous to those carried out by artificial neural networks. Moreover, we show that the chemical network can be trained \emph{in vitro} using directed evolution. In our framework, each molecular species functions as a “neuron,” and the dimensionless coefficients \(K_{ij}c_j \) play the role of “synaptic weights” connecting these neurons. Because these weights depend on both the concentrations \(c_i\) and the association constants \(K_{ij}\), they can, in principle, be experimentally tuned by adjusting the composition of the mixture or by modifying (or mutating) the molecules.

In analogy with DNNs, we designate: (i) a subset of \(n_{in}\) molecular concentrations \(\mathbf{c}_{\text{in}} = (c_1, \dots, c_{n_{in}})\) as the input layer; (ii) a non-overlapping subset of \(n_{out}\) fugacities \(\mathbf{f}_{\text{out}} = (f_{N-n_{out}+1}, \dots, f_N)\) as the output layer; and (iii) the remaining \(n_{h} = N - n_{in} - n_{out}\) molecules constitute a hidden layer (see Fig. \ref{fig:Fig1}a).

\rev1{These labels specify how species are controlled and read out rather than defining a feed-forward architecture. Apart from self-interactions, which are excluded separately, only input--input interactions are suppressed because the input molecules and their mutual interactions are treated as externally specified features of the chemical background being classified. Direct input--output interactions, as well as all pairwise interactions involving hidden or output species, are allowed. Output species are not chemically distinct from hidden species; they are designated as outputs only because their fugacities are measured and enter the loss function.}

Training the CDN involves adjusting both the concentrations (outside of the input layer) and the association constants to achieve a desired functionality. Once trained, the system’s output for any given input can be determined by measuring the output fugacities––for example, via Fluorescence Resonance Energy Transfer (FRET) labeling. Because the input is represented by a set of concentrations of species mixed together, we refer to it as an “input cocktail” (see Fig. \ref{fig:Fig1}b). Numerically, the simplest way to compute the model’s output is to iteratively solve the mass-action equilibrium equations \ref{eq:f} for \(\mathbf{f}\). \rev1{The present model is quasistatic: each input cocktail is applied to a fresh replica, allowed to equilibrate, and read once.}

Although our system bears conceptual similarities to DNNs, several important distinctions must be noted. First, the network structure of the CDN is inherently bidirectional––in contrast to the typical feedforward architecture of DNNs––since its connections, determined by the association constants, are naturally reciprocal. Second, the effective weights \(K_{ij}c_j\) in the CDN are strictly nonnegative. Third, the nonlinear activation function in the equilibrium equation, \(f(y) = \frac{1}{1+y}\), emerges directly from the law of mass action and may not be optimal for all learning tasks.

Notably, because our model corresponds to the minimization of a well-defined free energy, it shares similarities with Hopfield neural networks~\cite{Hopfield1982,Hopfield1984}, one of the simplest realizations of associative memory. \rev1{This analogy is limited to the presence of reciprocal couplings and a free-energy description. 
The uniqueness of the solution of the nonlinear transfer function in Eq.~\eqref{eq:f} follows directly from the convexity of the respective free energy \cite{Yan2008} \cite{varilly2012general}, making it impossible to store multiple stationary states in a model with given $K$ and ${\bf c}$.
}

\subsection{In vitro Training via Directed Evolution}

\rev1{The protocol described in this subsection is a proposed experimental route; all training results reported in this paper are numerical.}

While the generic description of CL in CDNs supports multiple molecular implementations, one of the simplest realizations is based on non-complementary DNA oligomers~\cite{Nikitin2023}. In this approach, the binding affinities \(K_{ij}\) are determined by the specific oligomer sequences and can be readily modulated via mutations.

We propose a concrete experimental implementation for the \emph{in vitro} training of such CDNs, as illustrated in Fig. \ref{fig:Fig1}c. In our design, the sequences corresponding to molecules in the hidden and output layers are concatenated into one or several \emph{master sequences} that serve as the system's genome. At each round of directed evolution, these master sequences are replicated with randomly induced mutations. Each master sequence is amplified using Polymerase Chain Reaction (PCR) and purified by removing complementary components generated by the PCR (e.g., using magnetic beads). Subsequently, the master sequences are cleaved into their constituent DNA oligomers (for instance, using restriction enzymes), diluted to desired concentrations $c_i$, and mixed with a cocktail of input oligomers at specified concentrations, \(\mathbf{c}_{\text{in}}\). The output fugacities, \(\mathbf{f}_{\text{out}}\), are then detected using techniques such as FRET.

\rev1{The master-sequence construction is a possible simplification rather than a requirement: cleavage sites can be placed in invariant linker regions outside the mutagenized binding domains, and multiple constructs or separately prepared oligomers are compatible with the same scheme. In the simulations, the dilution step is represented by the common hidden-layer concentration rescaling specified in the Appendix.}

\rev{Note that the system's ``genome'' consists exclusively of monomer sequences (e.g., ssDNA strands), and does not include dimers or other bound complexes. The genome encodes the association constants $K_{ij}$ through sequence-dependent binding affinities, rather than the instantaneous dynamical state of the network. The order in which sequences are concatenated carries no functional meaning beyond facilitating amplification and mutation.}

Training the system involves exposing several CDN variants to batches of input cocktails. These variants are ranked \emph{in silico} according to a loss function that quantifies the similarity between the measured and desired output fugacities. The best-performing variant is selected for the next round of \emph{in vitro} evolution (see Fig. \ref{fig:Fig1}c). Differences among variants arise from mutations that alter the association constants \(K_{ij}\) and, \rev1{in the numerical protocol, from rescaling the common hidden-layer concentration}. \rev1{A single best variant is selected from one loss evaluated over the complete batch and all output components; variants are not selected separately for different classes.}


In the present study, we model the effect of mutations as a biased multiplicative random walk of the association constants \(K_{ij}\):
\begin{eqnarray}
K_{ij}(t+1)= K_{ij}(t) e^{\sigma_{\text{mut}}\left(  \eta_{ij}(t) -\eta_0\right)}
\label{eq:mut}
\end{eqnarray}
Here, \(t\) denotes the round of in vitro evolution, and \(\eta_{ij}(t)\) is a Gaussian white noise with \(\langle \eta_{ij}(t) \rangle = 0\) and \(\langle \eta_{ij}(t) \eta_{ij}(t') \rangle = \delta_{tt'}\). The variance \(\sigma_{\text{mut}}^2\) characterizes the mutation rate and its impact on binding free energy, while the drift parameter \(\eta_0 > 0\) reflects the mutational bias toward weakening binding interactions. Note that (i) \(\ln K_{ij}\) is proportional to the binding free energy between molecules \(i\) and \(j\), which justifies our choice of a multiplicative random walk; and (ii) mutations are statistically more likely to weaken binding rather than strengthen it, rendering the random walk negatively biased. In a more detailed study, one may explicitly introduce random mutations into the sequences of individual DNA oligomers and calculate \(K_{ij}\) using the standard model described in \cite{santalucia1998unified-7e0}.

\subsection{Chemical Classifier}

While CDNs can, in principle, perform a wide variety of learning tasks, here we demonstrate their utility as a multiclass classifier. In our study, a certain number of different classes of input cocktails are generated randomly, as shown in Fig. \ref{fig:Fig2}a. Each cocktail is characterized by a set of input concentrations, \(\mathbf{c}_{\text{in}}\), drawn from a log-normal distribution. To emulate the noise present in real experiments, these input concentrations are further scrambled by multiplicative noise. 
For the outputs, one-hot encoding is employed so that the number of outputs, \(n_{out}\), equals the number of cocktail classes to be distinguished. As illustrated in Fig. \ref{fig:Fig2}b, one-hot encoding designates a specific output as active (high fugacity or “on”) for each class, with all other outputs remaining inactive (low fugacity or “off”).

At each step of in vitro evolution, we generate a balanced batch in which each class of patterns is represented by an equal number of examples and each example subject to a different realization of noise. Unlike traditional deep neural network training which typically involves gradient descent and back-propagation, our \rev1{numerical simulations} utilize an evolutionary learning process. In this process, variants are generated by (i) randomly and independently mutating all \rev1{allowed} association constants according to Eq. \ref{eq:mut}, and (ii) \rev1{modulating the common hidden-layer concentration} as illustrated in Fig. \ref{fig:Fig1}c. Experimentally, this setup can be realized by first amplifying batches with mutations introduced via random mutagenesis using PCR and then diluting them by preset factors. The winning variant, characterized by a specific combination of sequences and concentrations, is selected based on a custom loss function.

Since the goal of our training is to maximize the contrast between the two 
states of the output components, the loss function is designed to penalize poor contrast. \rev{A reasonable loss function would be the logarithm of the ratio between the output signal in the desired ``off'' and ``on'' states. However, in a training batch each output component is evaluated on multiple data points. To aggregate them, we employ the arithmetic averaging for the ``off'' samples and the geometric averaging for the ``on'' samples.
This hybrid type of averaging emphasizes the worst contrast ratio among multiple samples in a batch.
The overall loss is then taken as the worst contrast ratio across all output components:}
\begin{align}
    \Lambda_{\rm contrast} = \max_{i \in \rm out} 
    \left[ 
    \ln \langle f^{\alpha \downarrow}_i \rangle_{\alpha \downarrow} 
    - 
    \left\langle \ln f^{\alpha \uparrow}_i \right\rangle_{\alpha \uparrow} 
    \right]
    =  \nonumber \\
    = \max_{i \in \rm out} 
    \left[ 
    \ln \left(
    \frac{\langle (1-\phi_i^\alpha) f^{\alpha}_i \rangle_\alpha}
    {1-\langle \phi_i^\alpha \rangle_\alpha}
    \right) 
    - 
    \frac{\left\langle \phi_i^\alpha \ln f^{\alpha}_i \right\rangle_\alpha}
    {\langle \phi_i^\alpha \rangle_\alpha} 
    \right] ,
    \label{eq:loss_log1}
\end{align}

\rev{Here the index $\alpha$ labels samples in the training batch, and $f_i^\alpha$ denotes the output fugacity of component $i$ for sample $\alpha$. The binary variable $\phi_i^\alpha \in \{0,1\}$ specifies the desired target state of component $i$, with $\phi_i^\alpha=1$ and $0$ corresponding to the ``on'' and ``off'' states, respectively. The notation $\langle\cdot\rangle_{\alpha\uparrow}$ and $\langle\cdot\rangle_{\alpha\downarrow}$ denotes averages over the subsets of samples with $\phi_i^\alpha=1$ and $\phi_i^\alpha=0$, while $\langle\cdot\rangle_\alpha$ denotes an average over the full batch.}

\rev{The maximum over output components selects the least contrasted (worst-performing) output, enforcing uniform classification performance across all outputs. For the loss to be well defined, each output component must be represented by both ``on'' and ``off'' samples in the batch, i.e., $0 < \langle \phi_i^\alpha \rangle_\alpha < 1$.}

\rev1{For the balanced 48-sample batches used here, each output has 16 ``on'' samples (independent noisy realizations of its class) and 32 ``off'' samples; the geometric average is therefore taken over multiple samples, not over a single one.}

\section{Results}
\subsection{Performance and Efficiency of the Chemical Classifier}
\begin{figure}
    \centering    \includegraphics[width=\linewidth]{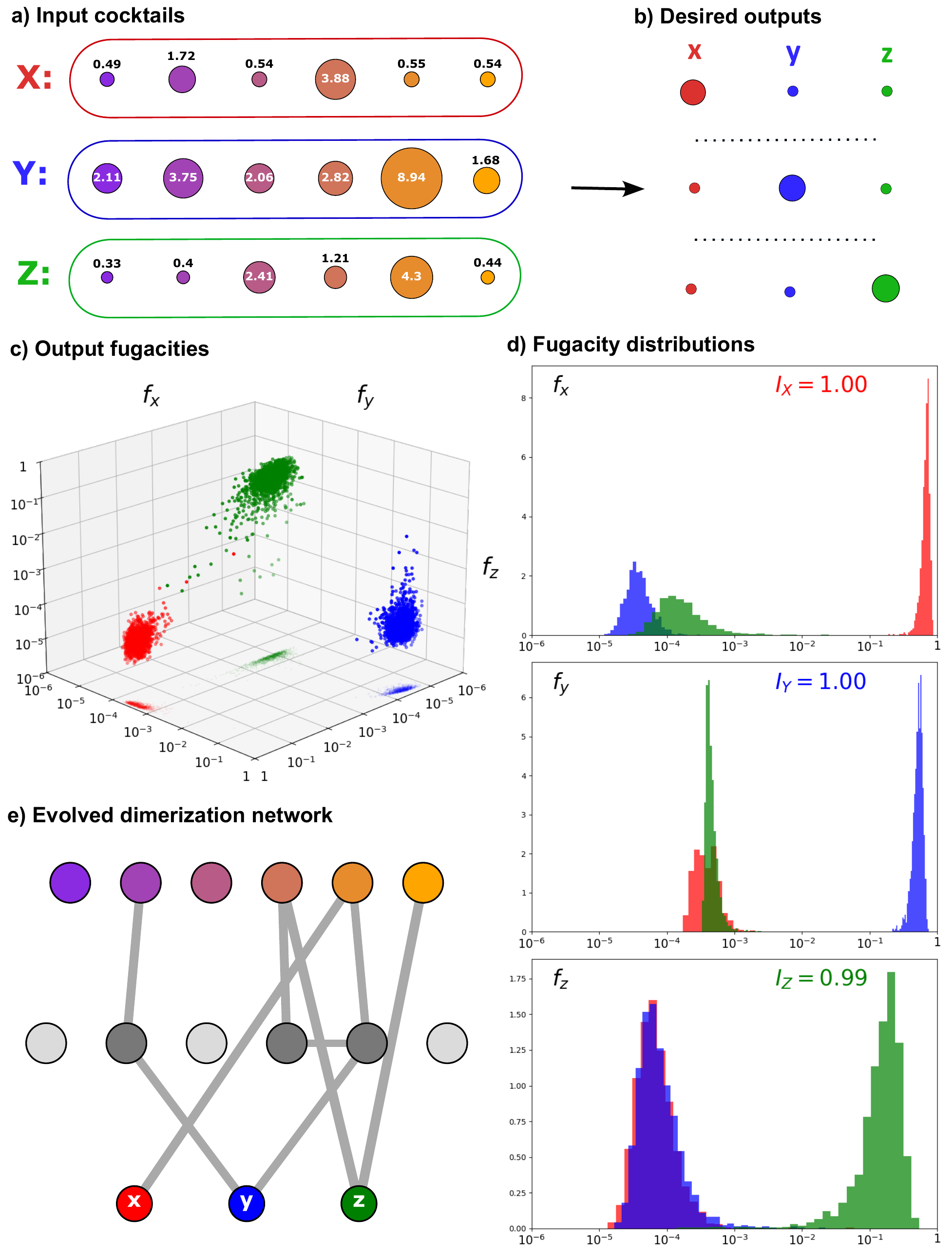}   
    \caption{Performance of the chemical classifier trained via directed evolution. (a) Randomly generated three input cocktails. (b) One-hot encoded target outputs for three cocktail classes (X, Y, Z). (c) 3D scatter plot showing class separation based on output fugacities at 20\% noise. Each point corresponds to a single noisy realization of inputs in the test dataset. The colors correspond to the true class of the cocktail. (d) Histograms of output fugacities plotted in panel c, showing strong contrast between “on” and “off” states. \rev{Different colors correspond to histograms of the same output component (e.g. $f_x$) for different input  classes: $X$ (red), $Y$ (blue), and $Z$ (green). The mutual information $I$ between inputs (e.g. $X$) and its corresponding output (e.g., $f_x$), is indicated for  each of the output components. } (e) Network diagram of the evolved CDN after 400 generations of in vitro evolution. Grayscale shading of hidden layer nodes indicates their overall connectivity. In this specific example, all displayed interactions have identical association constants, fixed at the upper saturation limit. 
    An interactive version of this figure, allowing scanning over multiple parameters and cocktail sets, is available in Ref.~[34].}
    \label{fig:Fig2}
\end{figure}

We numerically investigated a chemical classifier comprising \(n_{out}=3\) classes of cocktails (labeled \(X\), \(Y\), and \(Z\)) with \(n_{in}=6\) input channels. The input class concentrations $c_i$ were randomly generated from a log-normal distribution 
$c_0e^{\mathcal{N}(0, \sigma^2)}$ with $\sigma^2=2$. 
The actual concentrations presented to the chemical classifier were then scrambled by the multiplicative noise $c_i e^{\mathcal{N}(0, (\nu\sigma)^2)} $, emulating inevitable noise present in real experimental conditions. The parameter \(\nu\) thus quantifies the noise-to-signal (N/S) ratio in our inputs. Figure \ref{fig:Fig2} illustrates one realization of inputs generated with a N/S ratio set to  20\%. At this noise level, the evolutionarily trained CDN successfully discriminates among the three classes. This is evident in the 3D scatter plot (Fig. \ref{fig:Fig2}c), where data points form three distinct clusters—red for \(X\), blue for \(Y\), and green for \(Z\).

Furthermore, the fugacities of individual output components (\(x\), \(y\), and \(z\)) clearly resolve the classes, as demonstrated by the histograms in Fig. \ref{fig:Fig2}d. The contrast between the output fugacities of the “on” and “off” states spans three orders of magnitude. Note that all axes in Figs. \ref{fig:Fig2}c–d are displayed on a logarithmic scale.

Figure \ref{fig:Fig2}e presents the structure of the evolutionarily trained chemical network after 400 generations. In this network diagram, the thickness of each edge reflects the logarithm of the corresponding association constant \(K_{ij}\). For clarity, edges with \(K_{ij}<1\) have been omitted since their contributions to the classifier function are negligible. The strongest connection corresponds to \(c_0K_{ij}=10^4\), which is the upper cutoff set in our evolution procedure. \rev1{This cutoff represents the finite affinity range allowed by the model; increasing it would primarily increase the achievable on/off contrast, because the strongest selected links are driven toward the largest available affinity.}
\begin{figure}
    \centering
    \includegraphics[width=\linewidth]{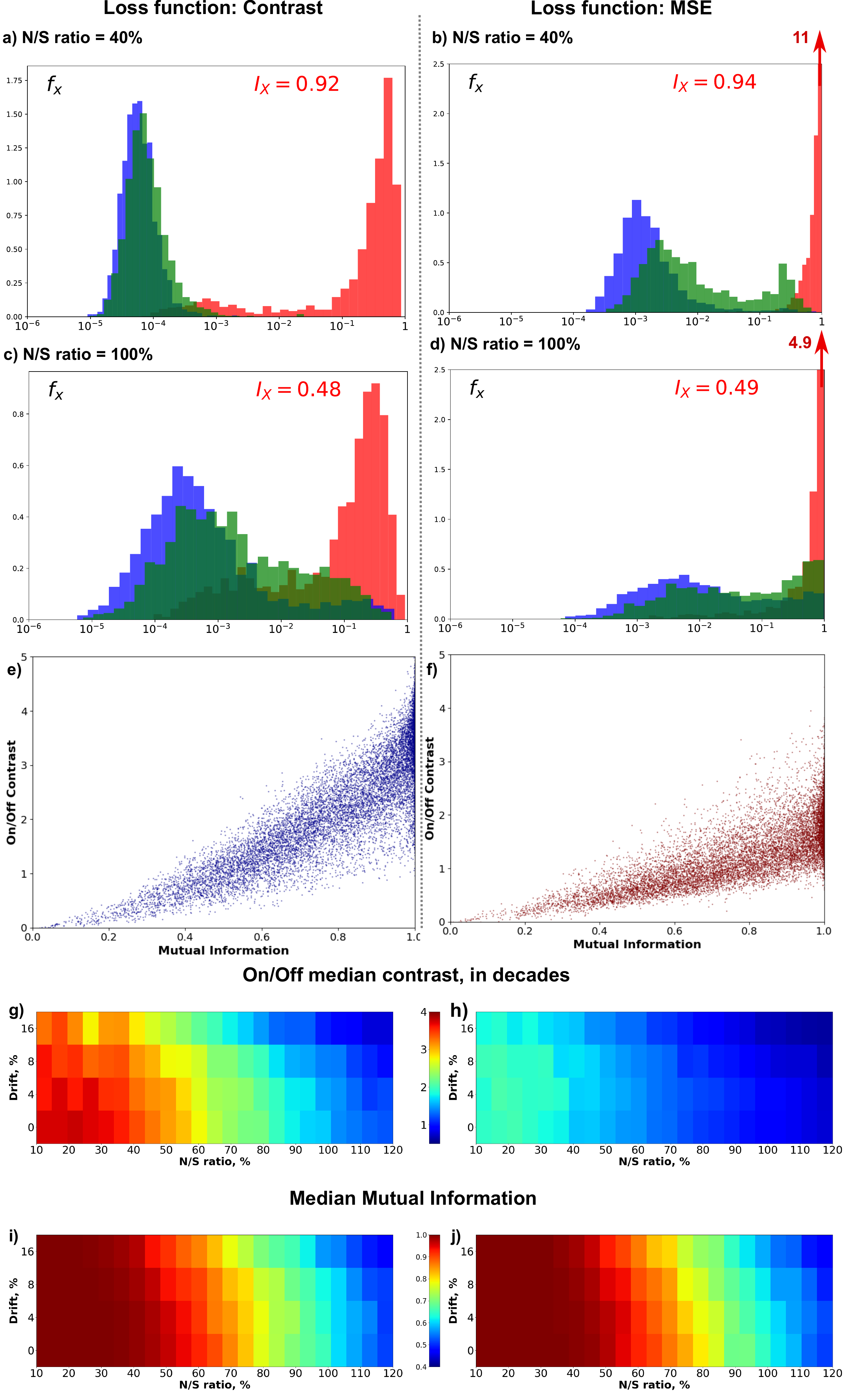}   
    \caption{Effect of noise and choice of loss function on classifier performance. Throughout the figure, the left column corresponds to the contrast loss, while the right column corresponds to the MSE loss. (a–d) Output histograms and mutual information at 40\% and 100\% noise-to-signal ratios, showing the effect of the loss function on output separation and information content. (e, f) Scatterplots illustrating the correlation between on/off contrast and mutual information for both loss functions. Points correspond to different  realizations of input classes, noise levels and drift parameters. (g, h) Heatmaps of on/off contrast as a function of drift and noise-to-signal ratio. (i, j) Corresponding heatmaps of mutual information, demonstrating the robustness of information transmission under varying training conditions.}
    \label{fig:Fig3}
\end{figure}

The efficiency of the chemical classifier is evaluated using two complementary metrics. The first measure, the on/off contrast, is defined as the logarithmic difference (expressed in decades) between the output fugacities in their "on" and "off" states. The second metric is the Mutual Information (MI) between the inputs and outputs, an information-theoretic quantity that generalizes the correlation coefficient. In this context, MI captures the degree of overlap between the "on" and "off" histograms of the outputs.

The mutual information between the binary variable \(\mathcal{X}\) (assigned a value of 1 for cocktails of class \(X\) and 0 otherwise) and the corresponding output fugacity \(f_x\) is given by:
\begin{align}
I_X = \frac{1}{S_0}\sum_{f_x}\Bigg[p_\mathcal{X}\,P(f_x|\mathcal{X})\ln\left(\frac{P(f_x|\mathcal{X})}{P(f_x)}\right) + \nonumber \\
(1-p_\mathcal{X})\,P(f_x|\bar{\mathcal{X}})\ln\left(\frac{P(f_x|\bar{\mathcal{X}})}{P(f_x)}\right)\Bigg]  \label{eq:MI}
\end{align}
Here, \(p_\mathcal{X}=1/n_{out}\) denotes the fraction of samples corresponding to class \(X\), and\[
P(f_x)=p_\mathcal{X}P(f_x|\mathcal{X}) + (1-p_\mathcal{X})P(f_x|\bar{\mathcal{X}})
\] is the total probability of observing a particular fugacity \(f_x\). This expression assumes a balanced testing set, with equal representation of each of the \(n_{out}\) classes. The normalization factor $ S_0 = -\left(p_\mathcal{X}\ln p_\mathcal{X} + (1-p_\mathcal{X})\ln(1-p_\mathcal{X})\right)$
is selected so that \(I_X=1\) corresponds to the maximal MI—achieved when the "on" and "off" histograms of \(f_x\) are completely non-overlapping—while \(I_X=0\) indicates statistical independence between \(\mathcal{X}\) and \(f_x\). Analogous definitions hold for the outputs \(y\) and \(z\), with the high fidelity of our chemical classifier evidenced by values \(I_X\), \(I_Y\), and \(I_Z\) all approaching 1 (Fig. \ref{fig:Fig2}d).

\subsection*{Effects of noise and training parameters}
Figure \ref{fig:Fig3} illustrates how both the noise level and the choice of loss function affect the efficiency of our classifier. As expected, an increase in the noise-to-signal (N/S) ratio, \(\nu\), results in a deterioration of performance (Figs. \ref{fig:Fig3}a–d). Specifically, comparing panels (a) and (b) for an N/S ratio of 40\% with panels (c) and (d) for 100\% clearly shows that MI drops markedly as noise increases, evident from the increased overlap between the "on" (red) and "off" (blue and green) histograms.

The selection of the loss function plays a critical role, particularly in terms of On/Off contrast. Our bespoke loss function, Eq. \ref{eq:loss_log1}, used for generating Figs. \ref{fig:Fig3}(a) and (c)—is specifically engineered to maximize contrast, yielding significantly better on/off separation than the conventional mean-squared error (MSE) loss function, which was employed in Figs. \ref{fig:Fig3}(b) and (d). While the MSE loss produces a sharper peak in the "on" state, it results in a weaker On/Off contrast. These findings indicate that the optimal choice of loss function should be guided by the specific requirements of the intended application. The role played by the choice of the loss function, and the connection between the two metrics used, is apparent from the comparison of two scatter plots, \ref{fig:Fig3}(e) and (f). These plots indicate a strong correlation between the Mutual Information, and On/Off contrast parameter, but the latter provides a valuable additional insight once MI saturates at its maximum value 1. As we have already observed in the specific example above, the use of MSE loss function results in significantly lower On/Off contrasts.   

Heatmaps in Figs. \ref{fig:Fig3}g–j further quantify these trends as functions of both the noise-to-signal (N/S) ratio (x-axis) and the negative drift parameter $\eta_0$ from the evolutionary training defined in Eq. \ref{eq:mut} (y-axis). Increases in either the noise-to-signal ratio or the drift parameter degrade classifier performance. Notably, while the loss functions \(\Lambda_{\text{contr}}\) (Fig. \ref{fig:Fig3}(g) and MSE (Fig. \ref{fig:Fig3}(h) exhibit dramatically different on/off contrasts, the mutual information remains virtually unchanged, as shown in Figs. \ref{fig:Fig3}(i)–(j).

\begin{figure}
    \centering
    \includegraphics[width=\linewidth]{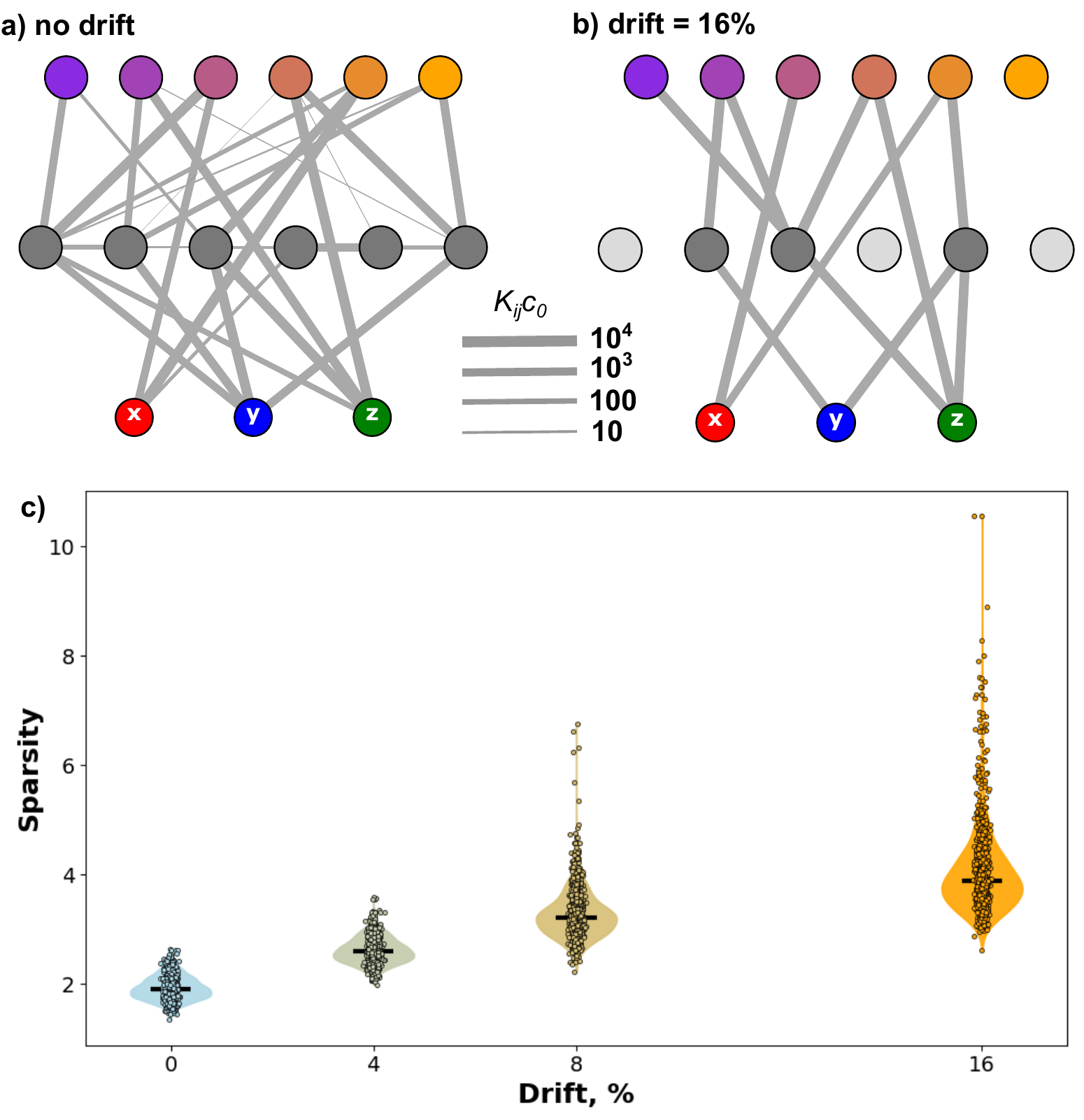}   
        \caption{Drift-induced sparsity in trained chemical networks. (a, b) Evolved CDN network topologies at zero drift (left) and 16\% drift (right), with edge thickness proportional to log-transformed binding strengths ($\log_{10} K_{ij}c_0$). Weak interactions ($K_{ij}c_0 < 1$) are omitted. Grayscale of hidden nodes reflects their connectivity, defined as the sum of their edge weights.  (c) Network sparsity  measured by the dispersion coefficient (standard deviation divided by mean) of edge weights as a function of the evolutionary drift.}
    \label{fig:Fig4}
\end{figure}

Increasing the drift parameter leads to a modest decrease in the chemical classifier’s noise tolerance while simultaneously rendering the trained dimerization network sparser. In other words, the network achieves its function through a smaller number of strong links, as demonstrated in Figs. \ref{fig:Fig4}ab. To construct these graphs, we disregarded all weak interactions satisfying $K_{ij}c_0<1$. The thickness of the remaining edges is proportional to $\ln K_{ij}c_0$, which, in turn, is proportional to the corresponding binding free energies. Specifically, the edge weights $w_{ij}$ are defined as

\begin{equation}
   w_{ij} = \begin{cases}
  \log_{10} (K_{ij}c_0)  & K_{ij}c_0 > 1 \\
  0 & \text{otherwise.}
  \end{cases}
\end{equation}

We quantify the sparsity of a weighted network in terms of the dispersion coefficient (the ratio of the standard deviation to the mean) of its edge weights $w_{ij}$, and we show the relationship between sparsity and the evolutionary drift parameter in Fig. \ref{fig:Fig4}c. The observed increase in sparsity with drift is expected, as the drift tends to weaken all bindings and eliminate redundant interactions. In this regard, the drift plays a role analogous to regularization in standard machine learning. Consequently, it is not surprising that the simpler, sparser networks exhibit lower noise tolerance compared to their more connected counterparts (see Figs. \ref{fig:Fig3}g–h).

\section{Discussion}
\subsection{Comparison between Evolutionary Learning and Gradient Descent}
\begin{figure}
\centering
\includegraphics[width=\linewidth]{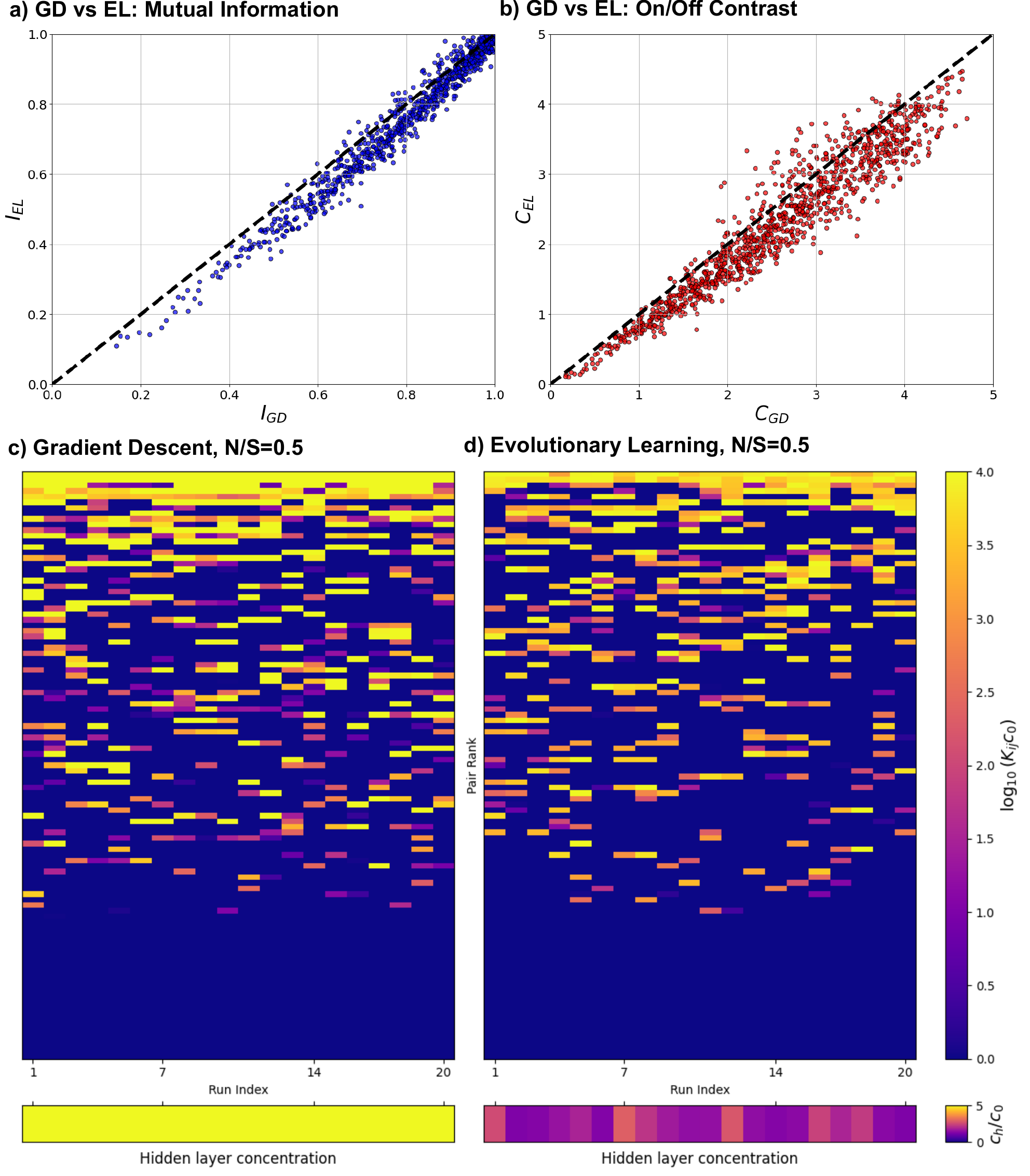}
\caption{Comparison of network structures obtained by evolutionary learning (EL) and gradient descent (GD).
\textbf{(a,b)} Scatter plots comparing the chemical classifier performance for EL and GD across multiple input cocktails and noise levels. Panel (a) shows mutual information, and panel (b) shows on/off contrast. Each point represents performance averaged over multiple drift (EL) and regularization (GD) parameters. The dashed diagonal indicates equal performance; points below the diagonal correspond to cases where GD outperforms EL.
\textbf{(c,d)} Heatmaps summarizing network variability across 20 independent training runs for GD (panel c) and EL (panel d), performed on the same cocktail set at $N/S=0.5$. The top sub-panels show edge weights $w_{ij}=\log_{10} (c_0K_{ij}), c_0K_{ij}>1$, ordered by their mean strength across all runs; the bottom sub-panels show hidden layer concentrations $c_h/c_0$. Columns correspond to individual training runs and are hierarchically clustered by network similarity. \rev1{The heatmap y-axis is the pair rank. All hidden species share the same concentration by construction, so the lower strip reports one common $c_h$ per run; the high performance obtained under this restriction is part of the result.}}
\label{fig:Fig5}
\end{figure}

All of the results discussed above were obtained using evolutionary training of the chemical classifier, a method that can be implemented in vitro. In contrast, a more traditional approach to training neural networks relies on backpropagation and Gradient Descent (GD). However, because our system is bidirectional, unlike standard feedforward neural networks, backpropagation cannot be applied. Nonetheless, GD remains feasible and is implemented as described in the Appendix. As noted earlier, the regularization coefficient in the GD algorithm serves a role analogous to the drift parameter in Evolutionary Learning (EL). It is therefore instructive to compare CDNs trained using EL and GD (see Fig. \ref{fig:Fig5}). Each point in the scatterplot represents a specific combination of input cocktail and noise level. Because there is no one-to-one mapping between the drift parameter and the regularization coefficient, each point reflects an average over multiple values of drift (for EL) and regularization (for GD). As shown in Fig. \ref{fig:Fig5}a-b, the performance of the two training methods is highly correlated and generally comparable. However, GD exhibits a slight advantage over EL, as indicated by the clustering of points below the diagonal dashed line in both panels. 

To further probe the distinction between GD and EL, we compared the network structures and their reproducibility under each training scheme. We performed 20 independent runs of both EL and GD on the same cocktail set at $N/S = 0.5$. Figures \ref{fig:Fig5}(c) and \ref{fig:Fig5}(d) present these results as heatmaps of edge weights $w_{ij}$ and hidden‐layer concentrations $c_h/c_0$. Rows correspond to 90 pairs of interacting molecules, ranked by their interaction strength $\langle w_{ij}\rangle$ averaged over all runs, while columns represent individual runs. 
Strikingly, none of the runs produced an identical network topology, underscoring a highly degenerate and rugged loss‐landscape. Despite this diversity, classifier performance remains consistent across runs. In GD, five of the strongest network links persist across most solutions, and the hidden‐layer concentration invariably reaches its maximum, $c_h=5c_0$. In contrast, EL retains only two invariant links and exhibits considerable run‐to‐run variability in $c_h$
, which never attains the upper bound. From this perspective, EL behaves as though sampling at a higher effective temperature compared to GD. Remarkably, despite these significant structural differences, the final classifier performance under both methods remains nearly identical (Fig. \ref{fig:Fig5}a–b).

\subsection{Experimental feasibility and roadmap}

Implementing full evolutionary learning \emph{in vitro} is a long–term objective. Crucially, the CDN framework enables a staged path that yields meaningful experimental milestones well before a complete closed loop is realized.

\textbf{Stage 1: In-silico design \& direct implementation.} 
A functional classifier can first be trained entirely \emph{in silico}, either by gradient descent (GD) or virtual evolutionary learning (EL). The resulting target affinities $K_{ij}$ are then mapped to DNA sequences using established nearest-neighbor thermodynamics (SantaLucia) and implemented on a partial-complementarity platform already demonstrated for programmable, many-to-many DNA interactions \cite{Nikitin2023,santalucia1998unified-7e0}. This yields a fixed (non-evolving) CDN that should already exhibit robust one-hot classification, providing an immediate experimental validation.

\textbf{Stage 2: Small-scale directed evolution for refinement.}
Starting from a predesigned CDN, limited-cycle \emph{in vitro} evolution can fine-tune $K_{ij}$ and concentrations, leveraging standard selection toolkits (e.g., droplet microfluidics with FRET-based gating, bead capture, or enrichment based on Systematic Evolution of Ligands by Exponential Enrichment (SELEX)). Our simulations use 400 iterations to fully explore parameter space; however, the classifier’s performance typically approaches its plateau within $\lesssim$100 iterations (see Appendix E, Figs.~6--9), suggesting that early demonstrations can target tens of rounds. With basic automation, a round comprising amplification, cleanup/processing, mixing, readout, and selection can be completed in hours, and, importantly, executed in parallel across $10^2$–$10^3$ variants per batch. Thus, practical efforts can emphasize breadth (parallelism) over depth (very long runs), while still capturing most of the attainable function.

In sum, a pragmatic experimental program begins with \emph{in silico} training and direct sequence design, followed by modest-cycle evolutionary refinement on the same physical platform. This staged approach reduces risk and cost while steadily moving toward fully autonomous evolutionary chemical learning.

\section{Conclusions}
In this study, we developed and explored the concept of chemical learning in competitive dimerization networks (CDNs) and \rev1{demonstrated numerically a route to their \emph{in vitro} training} through directed evolution. Our results establish that CDNs, which harness intrinsic biomolecular interactions, are capable of performing complex classification tasks with high fidelity and robust noise tolerance.
\rev{We emphasize that the present work is computational and theoretical in nature. While all proposed protocols are experimentally feasible and grounded in existing DNA-based platforms, no experimental data are reported here. Instead, our goal is to establish a quantitative framework and a realistic roadmap for future experimental realizations of chemical learning.}

We found that our chemical classifier can reliably discriminate among various input cocktails by employing a one-hot encoding strategy. In this scheme, each cocktail triggers the activation of a specific output component by transitioning it to a free, unbound state, while all other outputs remain predominantly in a tightly bound state.  
\rev{Importantly, computation in CDNs is an emergent collective property of the entire network rather than a function of individual molecular components. Learning arises from the global reorganization of interaction strengths and concentrations, connecting our framework to broader themes in collective computation, self-organization, and distributed physical intelligence.}

A key aspect of our work is the \rev1{proposed use of \emph{in vitro}} training via directed evolution. In contrast to traditional deep learning approaches that depend on gradient descent and backpropagation, our protocol enables the network to iteratively evolve toward the desired functionality through selective pressure. This evolutionary process not only optimizes the association constants and concentrations within the network but also naturally regularizes the system by progressively eliminating redundant interactions, leading to sparser and more interpretable networks.

Furthermore, the possibility of a hybrid training strategy is particularly promising. An initial in silico phase can be used to design an approximate optimal dimerization network by determining target association constants through computational simulations. These computationally derived parameters can then be implemented in the physical system by engineering appropriate intermolecular interactions—such as selecting the right sequences of DNA, RNA, or proteins—and subsequently refined with in vitro directed evolution. This approach effectively marries the speed and flexibility of digital simulation with the adaptability and robustness intrinsic to experimental evolution.

A key advantage of evolutionary training is its adaptability in the absence of complete knowledge of the underlying system. A natural hybrid protocol combines \emph{in silico} pretraining by gradient descent with \emph{in vitro} evolutionary fine-tuning. This strategy is particularly valuable when binding parameters are unknown or drift over time because of changes in pH, ionic strength, or the presence of competing molecules. In this sense, adaptability is an intrinsic feature of chemical learning rather than a secondary benefit. \rev1{More recently, following the appearance of an earlier version of this manuscript, Ref.~\cite{Trifonova2025Trainable} demonstrated that highly connected chemical networks can be trained to perform a variety of tasks by adjusting only hidden-layer concentrations while keeping the pairwise affinities fixed. Since evolutionary learning can proceed without complete knowledge of the underlying interaction parameters, this result suggests an interesting simplified \emph{in vitro} protocol based solely on controlled dilution, without requiring a mutation step.}

The broader implications of our work extend well beyond the proof-of-concept demonstrated here. High-fidelity CDNs may find applications across a range of fields, from biomedical diagnostics, where they could help distinguish between healthy and diseased cellular states, to nanotechnology, in which adaptive chemical networks might mediate environment-responsive structural transformations. These potential applications highlight the versatility of chemical networks as a platform for unconventional information processing. \rev{Notably, the introduction of the contrast loss function is essential for achieving extreme signal separation: in optimized networks, ``on'' and ``off'' output fugacities differ by up to four orders of magnitude. Such high contrast is critical for practical applications, including molecular sensing and therapeutic activation, where false positives can have severe consequences.}

Overall, our study underscores the promise of chemical learning as an innovative paradigm for computation. By leveraging the laws of thermodynamics and kinetics inherent to biomolecular systems, CDNs offer a novel approach to information processing that sidesteps the need for conventional digital hardware. Inspired by evolution and intermolecular interactions in biological systems, this work lays the foundation for future advances in adaptive and energy-efficient molecular computing.

This research used the  Theory and Computation facility of the Center for Functional Nanomaterials (CFN), which is a U.S. Department of Energy Office of Science User Facility, at Brookhaven National Laboratory under Contract No. DE-SC0012704.
Computational resources have also been provided by the Consortium des \'Equipements de Calcul Intensif (CECI), funded by the Fonds de la Recherche Scientifique de Belgique (F.R.S.-FNRS) under Grant No. 2.5020.11 and by the Walloon Region. \rev{Research by S.M. was supported in part by grants from the NSF (DMS-2235451) and Simons Foundation (MPS-NITMB-31700005320) to the NSF-Simons National Institute for Theory and Mathematics in Biology (NITMB).} Codes and data generated as a part of  this work are available via GitHub repository \cite{Tkachenko2025github}.

\bibliography{main}

@article{phys_intelligence,
title = {Physical intelligence as a new paradigm},
journal = {Extreme Mechanics Letters},
volume = {46},
pages = {101340},
year = {2021},
issn = {2352-4316},
doi = {https://doi.org/10.1016/j.eml.2021.101340},
url = {https://www.sciencedirect.com/science/article/pii/S2352431621001012},
author = {Metin Sitti},
}

@book{Goodfellow2016,
  title        = {Deep Learning},
  author       = {Goodfellow, Ian and Bengio, Yoshua and Courville, Aaron},
  publisher    = {MIT Press},
  year         = {2016}
}

@article{krizhevsky2012imagenet,
author = {Krizhevsky, Alex and Sutskever, Ilya and Hinton, Geoffrey E.},
title = {ImageNet classification with deep convolutional neural networks},
year = {2017},
issue_date = {June 2017},
publisher = {Association for Computing Machinery},
address = {New York, NY, USA},
volume = {60},
number = {6},
issn = {0001-0782},
url = {https://doi.org/10.1145/3065386},
doi = {10.1145/3065386},
journal = {Commun. ACM},
month = may,
pages = {84–90},
numpages = {7}
}

@article{Liu2021PRX,
  title = {Supervised Learning in Physical Networks: From Machine Learning to Learning Machines},
  author = {Stern, Menachem and Hexner, Daniel and Rocks, Jason W. and Liu, Andrea J.},
  journal = {Phys. Rev. X},
  volume = {11},
  issue = {2},
  pages = {021045},
  numpages = {18},
  year = {2021},
  month = {May},
  publisher = {American Physical Society},
  doi = {10.1103/PhysRevX.11.021045},
  url = {https://link.aps.org/doi/10.1103/PhysRevX.11.021045}
}

@article{optical,
author = {Xing Lin  and Yair Rivenson  and Nezih T. Yardimci  and Muhammed Veli  and Yi Luo  and Mona Jarrahi  and Aydogan Ozcan },
title = {All-optical machine learning using diffractive deep neural networks},
journal = {Science},
volume = {361},
number = {6406},
pages = {1004-1008},
year = {2018},
doi = {10.1126/science.aat8084},
}

@article{memristors,
  title={Efficient and self-adaptive in-situ learning in multilayer memristor neural networks},
  author={Li, Can and Belkin, Daniel and Li, Yunning and Yan, Peng and Hu, Miao and Ge, Ning and Jiang, Hao and Montgomery, Eric and Lin, Peng and Wang, Zhongrui and others},
  journal={Nature communications},
  volume={9},
  number={1},
  pages={2385},
  year={2018},
  publisher={Nature Publishing Group UK London}
}

@article{photonics,
  title = {Photonics for artificial intelligence and neuromorphic computing},
  volume = {15},
  ISSN = {1749-4893},
  url = {http://dx.doi.org/10.1038/s41566-020-00754-y},
  DOI = {10.1038/s41566-020-00754-y},
  number = {2},
  journal = {Nature Photonics},
  publisher = {Springer Science and Business Media LLC},
  author = {Shastri,  Bhavin J. and Tait,  Alexander N. and Ferreira de Lima,  T. and Pernice,  Wolfram H. P. and Bhaskaran,  Harish and Wright,  C. D. and Prucnal,  Paul R.},
  year = {2021},
  month = jan,
  pages = {102–114}
}

@article{santalucia1998unified-7e0, 
  year     = {1998}, 
  title    = {A unified view of polymer, dumbbell, and oligonucleotide {DNA} nearest-neighbor thermodynamics}, 
  author   = {{SantaLucia}, John}, 
  journal  = {Proceedings of the National Academy of Sciences}, 
  issn     = {0027-8424}, 
  doi      = {10.1073/pnas.95.4.1460}, 
  pmid     = {9465037}, 
  pmcid    = {{PMC}19045}, 
  pages    = {1460--1465}, 
  number   = {4}, 
  volume   = {95}
}

@article{kramer2022multimodal,
  title={Multimodal perception links cellular state to decision-making in single cells},
  author={Kramer, Bernhard A and Sarabia Del Castillo, Jacobo and Pelkmans, Lucas},
  journal={Science},
  volume={377},
  number={6606},
  pages={642--648},
  year={2022},
  doi={10.1126/science.abf4062},
  publisher={American Association for the Advancement of Science}
}

@article{bhalla1999emergent,
  title={Emergent properties of networks of biological signaling pathways},
  author={Bhalla, Upinder S and Iyengar, Ravi},
  journal={Science},
  volume={283},
  number={5400},
  pages={381--387},
  year={1999},
  doi={10.1126/science.283.5400.381}
}

@article{bray1995protein,
  title={Protein molecules as computational elements in living cells},
  author={Bray, Dennis},
  journal={Nature},
  volume={376},
  number={6538},
  pages={307--312},
  year={1995},
  doi={10.1038/376307a0}
}

@inbook{poole2017dna-860,
  title = {Chemical Boltzmann Machines},
  ISBN = {9783319667997},
  ISSN = {1611-3349},
  url = {http://dx.doi.org/10.1007/978-3-319-66799-7_14},
  DOI = {10.1007/978-3-319-66799-7_14},
  booktitle = {DNA Computing and Molecular Programming},
  publisher = {Springer International Publishing},
  author = {Poole,  William and Ortiz-Muñoz,  Andrés and Behera,  Abhishek and Jones,  Nick S. and Ouldridge,  Thomas E. and Winfree,  Erik and Gopalkrishnan,  Manoj},
  year = {2017},
  pages = {210–231}
}

@article{brannetti2025covalent-c8c, 
  year     = {2025}, 
  title    = {Covalent Dynamic {DNA} Networks to Translate Multiple Inputs into Programmable Outputs}, 
  author   = {Brannetti, Simone and Gentile, Serena and Grosso, Erica Del and Otto, Sijbren and Ricci, Francesco}, 
  journal  = {Journal of the American Chemical Society}, 
  issn     = {0002-7863}, 
  doi      = {10.1021/jacs.4c13854}, 
  pmid     = {39905964}, 
  pmcid    = {{PMC}11848822}, 
  pages    = {5755--5763}, 
  number   = {7}, 
  volume   = {147}
}

@article{parresgold2025contextual-aac, 
  title = {Contextual computation by competitive protein dimerization networks},
  volume = {188},
  ISSN = {0092-8674},
  url = {http://dx.doi.org/10.1016/j.cell.2025.01.036},
  DOI = {10.1016/j.cell.2025.01.036},
  number = {7},
  journal = {Cell},
  publisher = {Elsevier BV},
  author = {Parres-Gold,  Jacob and Levine,  Matthew and Emert,  Benjamin and Stuart,  Andrew and Elowitz,  Michael B.},
  year = {2025},
  month = apr,
  pages = {1984--2002.e17}
}

@article{Nikitin2023, 
  year     = {2023}, 
  title    = {Non-complementary strand commutation as a fundamental alternative for information processing by {DNA} and gene regulation}, 
  author   = {Nikitin, Maxim P.}, 
  journal  = {Nature Chemistry}, 
  issn     = {1755-4330}, 
  doi      = {10.1038/s41557-022-01111-y}, 
  pmid     = {36604607}, 
  pages    = {70--82}, 
  number   = {1}, 
  volume   = {15}
}

@article{Maslov2007, 
  year     = {2007}, 
  title    = {Propagation of large concentration changes in reversible protein-binding networks}, 
  author   = {Maslov, S. and Ispolatov, I.}, 
  journal  = {Proc Natl Acad Sci U S A}, 
  issn     = {0027-8424}, 
  doi      = {10.1073/pnas.0702905104}, 
  pmid     = {17699619}, 
  pmcid    = {{PMC}1959437}, 
  eprint   = {0708.2421}, 
  url      = {https://www.ncbi.nlm.nih.gov/pubmed/17699619}, 
  pages    = {13655--60}, 
  number   = {34}, 
  volume   = {104}, 
}

@article{Yan2008, 
  year     = {2008}, 
  title    = {Fluctuations in mass-action equilibrium of protein binding networks}, 
  author   = {Yan, K. K. and Walker, D. and Maslov, S.}, 
  journal  = {Phys Rev Lett}, 
  issn     = {0031-9007}, 
  doi      = {10.1103/physrevlett.101.268102}, 
  pmid     = {19437675}, 
  eprint   = {0803.1471}, 
  url      = {https://www.ncbi.nlm.nih.gov/pubmed/19437675}, 
  pages    = {268102}, 
  number   = {26}, 
  volume   = {101}, 
}

@article{Maslov2007njp, 
  year     = {2007}, 
  title    = {Spreading out of perturbations in reversible reaction networks}, 
  author   = {Maslov, S. and Sneppen, K. and Ispolatov, I.}, 
  journal  = {New J Phys}, 
  issn     = {1367-2630}, 
  doi      = {10.1088/1367-2630/9/8/273}, 
  pmid     = {18046464}, 
  pmcid    = {{PMC}2098699}, 
  eprint   = {q-bio/0611026}, 
  url      = {https://www.ncbi.nlm.nih.gov/pubmed/18046464}, 

  pages    = {273}, 
  number   = {8}, 
  volume   = {9}, 
}

@article{Arvind2023,
   author = "Stern, Menachem and Murugan, Arvind",
   title = "Learning Without Neurons in Physical Systems", 
   journal= "Annual Review of Condensed Matter Physics",
   year = "2023",
   volume = "14",
   number = "Volume 14, 2023",
   pages = "417-441",
   doi = "https://doi.org/10.1146/annurev-conmatphys-040821-113439",
   url = "https://www.annualreviews.org/content/journals/10.1146/annurev-conmatphys-040821-113439",
   publisher = "Annual Reviews",
   issn = "1947-5462",
   type = "Journal Article",
  }

@article{Arvind2015,
author = {Arvind Murugan  and Zorana Zeravcic  and Michael P. Brenner  and Stanislas Leibler },
title = {Multifarious assembly mixtures: Systems allowing retrieval of diverse stored structures},
journal = {Proceedings of the National Academy of Sciences},
volume = {112},
number = {1},
pages = {54-59},
year = {2015},
doi = {10.1073/pnas.1413941112},
URL = {https://www.pnas.org/doi/abs/10.1073/pnas.1413941112},
eprint = {https://www.pnas.org/doi/pdf/10.1073/pnas.1413941112},
}

@article{Hopfield1982,
  title = {Neural networks and physical systems with emergent collective computational abilities.},
  volume = {79},
  ISSN = {1091-6490},
  url = {http://dx.doi.org/10.1073/pnas.79.8.2554},
  DOI = {10.1073/pnas.79.8.2554},
  number = {8},
  journal = {Proceedings of the National Academy of Sciences},
  publisher = {Proceedings of the National Academy of Sciences},
  author = {Hopfield,  J J},
  year = {1982},
  month = apr,
  pages = {2554–2558}
}

@article{Hopfield1984,
  title = {Neurons with graded response have collective computational properties like those of two-state neurons.},
  volume = {81},
  ISSN = {1091-6490},
  url = {http://dx.doi.org/10.1073/pnas.81.10.3088},
  DOI = {10.1073/pnas.81.10.3088},
  number = {10},
  journal = {Proceedings of the National Academy of Sciences},
  publisher = {Proceedings of the National Academy of Sciences},
  author = {Hopfield,  J J},
  year = {1984},
  month = may,
  pages = {3088–3092}
}

@article{Klumpe2023,
  title = {The computational capabilities of many-to-many protein interaction networks},
  volume = {14},
  ISSN = {2405-4712},
  url = {http://dx.doi.org/10.1016/j.cels.2023.05.001},
  DOI = {10.1016/j.cels.2023.05.001},
  number = {6},
  journal = {Cell Systems},
  publisher = {Elsevier BV},
  author = {Klumpe,  Heidi E. and Garcia-Ojalvo,  Jordi and Elowitz,  Michael B. and Antebi,  Yaron E.},
  year = {2023},
  month = jun,
  pages = {430–446}
}

@article{Adam,
  title={Adam: A method for stochastic optimization},
  author={Kingma, Diederik P},
  journal={Preprint:},
volume ={arXiv},
pages ={1412.6980},
  year={2014}
}

@misc{Suri2024,
  author = {Floyd,  Carlos and Dinner,  Aaron R. and Murugan,  Arvind and Vaikuntanathan,  Suriyanarayanan},
  title = {Limits on the computational expressivity of non-equilibrium biophysical processes},
   journal={Preprint:},
volume ={arXiv},
pages ={2409.05827},
  year = {2024},
}

@article{Wright2022,
  title = {Deep physical neural networks trained with backpropagation},
  volume = {601},
  ISSN = {1476-4687},
  url = {http://dx.doi.org/10.1038/s41586-021-04223-6},
  DOI = {10.1038/s41586-021-04223-6},
  number = {7894},
  journal = {Nature},
  publisher = {Springer Science and Business Media LLC},
  author = {Wright,  Logan G. and Onodera,  Tatsuhiro and Stein,  Martin M. and Wang,  Tianyu and Schachter,  Darren T. and Hu,  Zoey and McMahon,  Peter L.},
  year = {2022},
  month = jan,
  pages = {549–555}
}

@article{Science2023,
author = {Ali Momeni  and Babak Rahmani  and Matthieu Malléjac  and Philipp del Hougne  and Romain Fleury },
title = {Backpropagation-free training of deep physical neural networks},
journal = {Science},
volume = {382},
number = {6676},
pages = {1297-1303},
year = {2023},
}

@article{Adleman1994,
  author  = {Adleman, Leonard M.},
  title   = {Molecular computation of solutions to combinatorial problems},
  journal = {Science},
  year    = {1994},
  volume  = {266},
  number  = {5187},
  pages   = {1021--1024},
  doi     = {10.1126/science.7973651}
}

@article{Seelig2006,
  author  = {Seelig, Georg and Soloveichik, David and Zhang, David Y. and Winfree, Erik},
  title   = {Enzyme-free nucleic acid logic circuits},
  journal = {Science},
  year    = {2006},
  volume  = {314},
  number  = {5805},
  pages   = {1585--1588},
  doi     = {10.1126/science.1132493}
}

@article{Soloveichik2010,
  author  = {Soloveichik, David and Seelig, Georg and Winfree, Erik},
  title   = {DNA as a universal substrate for chemical kinetics},
  journal = {Proceedings of the National Academy of Sciences},
  year    = {2010},
  volume  = {107},
  number  = {12},
  pages   = {5393--5398},
  doi     = {10.1073/pnas.0909380107}
}

@article{Srinivas2017,
  author  = {Srinivas, Niranjan and Parkin, James and Seelig, Georg and Winfree, Erik and Soloveichik, David},
  title   = {Enzyme-free nucleic acid dynamical systems},
  journal = {Science},
  year    = {2017},
  volume  = {358},
  number  = {6369},
  pages   = {eaal2052},
  doi     = {10.1126/science.aal2052}
}

@misc{Tkachenko2025github,
  author       = {Alexei Tkachenko},
  title        = {ChemClassifierWeb: Data and Code Repository},
  year         = {2025},
  howpublished = {\url{https://github.com/atkachen00/ChemClassifierWeb}},
  note         = {Data and code available at the provided URL}
}

@article{Stanley2019Neuroevolution,
  author  = {Stanley, Kenneth O. and Clune, Jeff and Lehman, Joel and Miikkulainen, Risto},
  title   = {Designing Neural Networks through Neuroevolution},
  journal = {Nature Machine Intelligence},
  volume  = {1},
  pages   = {24--35},
  year    = {2019},
  doi     = {10.1038/s42256-018-0006-z}
}

@article{Wang2021DirectedEvolution,
  author  = {Wang, Yajie and Xue, Pu and Cao, Mingfeng and Yu, Tianhao and Lane, Stephan T. and Zhao, Huimin},
  title   = {Directed Evolution: Methodologies and Applications},
  journal = {Chemical Reviews},
  volume  = {121},
  number  = {20},
  pages   = {12384--12444},
  year    = {2021},
  doi     = {10.1021/acs.chemrev.1c00260}
}

@article{Trifonova2025Trainable,
  author  = {Trifonova, Kristina and Falk, Martin J. and Rouches, Mason and Vaikuntanathan, Suriyanarayanan and Elowitz, Michael B. and Murugan, Arvind},
  title   = {Trainable Computation in Molecular Networks},
  journal = {bioRxiv},
  year    = {2025},
  doi     = {10.64898/2025.12.28.696421},
  note    = {Preprint; first posted December 30, 2025; version 2 posted May 27, 2026}
}

@article{varilly2012general,
  title={A general theory of DNA-mediated and other valence-limited colloidal interactions},
  author={Varilly, Patrick and Angioletti-Uberti, Stefano and Mognetti, Bortolo M and Frenkel, Daan},
  journal={The Journal of chemical physics},
  volume={137},
  number={9},
  year={2012},
  publisher={AIP Publishing}
}

\pagebreak

\newpage
\renewcommand{\thefigure}{S\arabic{figure}}
\setcounter{figure}{0}
\renewcommand{\theequation}{S\arabic{equation}}
\setcounter{equation}{0}

\section*{Appendix}

\subsection{Loss functions}
In our modeling of evolutionary training we employed two different loss functions, one of which is the "contrast" loss  introduced by Eq.(\ref{eq:loss_log1}) in the main text, the other is the 
standard Mean Square Error (MSE) between the desired and observed output fugacities:
\begin{align}
\Lambda_{\rm contrast}& = \max_{i \in \rm out} \left[ \ln \left(\frac{\langle (1-\phi_i^\alpha) f^{\alpha}_i \rangle_\alpha}{1-\langle \phi_i^\alpha \rangle_\alpha}\right) - \frac{\left\langle \phi_i^\alpha \ln f^{\alpha}_i \right\rangle_\alpha}{\langle \phi_i^\alpha \rangle_\alpha} \right]
\label{loss:contrast}\\
\Lambda_{\rm mse} &= \left\langle \left(f_{i}^\alpha- \phi_i^\alpha\right)^2\right\rangle_{\alpha, i} 
\label{loss:mse}
\end{align}
\subsection{Performance metrics} The performance of the chemical classified was characterized by using the Mutual Information, Eq. \ref{eq:MI}, and on/off Contrast:
\begin{align}
D_x=\langle\log_{10}f_x |\mathcal{X}\rangle -\max\left[\langle\log_{10}f_x |\mathcal{Y}\rangle, \langle\log_{10}f_x |\mathcal{Z}\rangle\right] 
\end{align}
Here $\mathcal{X}$, $\mathcal{Y}$, $\mathcal{Z}$ are binary variables that indicate that the input belongs to a specific class. The contrast measures for $y$ and $z$ outputs are defined analogously.

\subsection{Evolutionary Training}

\rev{At each training and testing step, the fugacities are obtained by numerically solving Eq.~(2) for the given set of association constants, concentrations, and input cocktails. These equilibrium fugacities serve as the model outputs used both for loss evaluation during training and for performance assessment during testing.}

At each iteration step, we begin with the current best-performing system and generate $M = 49$ of its mutants by evolving the association matrix $K_{ij}$ according to Eq.\ (\ref{eq:mut}), with $\sigma_{\text{mut}} = 0.5$. For each of these $M$ mutants, as well as for the original system, we generate three variants by scaling the concentrations of the hidden layer components by factors $1$, $1 + \Delta$, and $1/(1 + \Delta)$, with $\Delta=0.015$. To ensure physical plausibility, the evolving association constants and concentrations are constrained by caps: $c_{\min} \leq c \leq c_{\max}$ ($c_{\min}=0.2$, $c_{\max}=5.0$), and $K_{ij} \leq K_{\max} = 10^4 / c_0$.

\rev1{Self-interactions and input--input interactions are held at zero; all other pairwise association constants are allowed to evolve. Only the common hidden-layer concentration is rescaled in this step, while output concentrations are fixed.}

Our procedure results in a total of $3(1 + M) = 150$ variants that are then evaluated against batches of 48 input cocktails, consisting of 16 cocktails for each of the $n_{\text{out}} = 3$ output classes. The input concentrations are perturbed by multiplicative log-normal noise: $c_i \, e^{\mathcal{N}(0, (\nu \sigma)^2)}$. The variant that minimizes the chosen loss function is selected as the new best-performing system, and the procedure is repeated. \rev1{Thus one global winner is selected from its performance over all three classes.}

\subsection{Gradient Descent}
We start by rewriting Eq.(\ref{eq:f}) in terms of free concentrations (activities) of the components, $x_i=c_if_i$:
\begin{equation}
x_i=\frac{c_i}{1+\sum_jK_{ij}x_j}
\end{equation}
To evaluate gradients of $\Lambda$ with respect to the model parameters, we expand this equation in terms of small increments of concentrations, activities and association constants: 
\begin{equation}
\delta x_i=\frac{x_i\delta c_i}{c_i} - \frac{x_i^2}{c_i}\sum_j\left(K_{ij} \delta x_j + x_j\delta K_{ij}\right)
\end{equation}
This set of  equations can be rewritten as
\begin{equation}
\left(\widehat K +\widehat T^{-1}\right)\delta {\bf x}= \left({\widehat X}^{-1} \delta {\bf c}-\delta {\widehat K}{\bf x}  \right)
\end{equation}
Here  $T_{ij}=\delta_{ij}x_i^2/c_i$ and $X_{ij}= \delta_{ij}x_i$ 
 This result allows one to calculate the gradients of the loss functions, Eqs.\ (\ref{loss:contrast}), (\ref{loss:mse}), in the space of  model parameters, ${\widehat K}$ and $\bf c$:  
\begin{align}
\frac{\partial \Lambda}{\partial c_i } &=\left\langle\frac{\partial \Lambda}{ \partial {\bf x}^{\alpha}_{\rm out}}\frac{\partial {\bf x}^\alpha_{\rm out}}{\partial c_i} \right\rangle_\alpha =\left\langle\frac{F_i^\alpha}{x_i^\alpha}\right\rangle_\alpha \\
\frac{\partial \Lambda}{\partial{\widehat K} }&=-\left\langle {\bf F}^\alpha\otimes {\bf x}^\alpha +{\bf x}^\alpha\otimes {\bf F}^\alpha -\widehat X^\alpha\widehat \Phi^\alpha\right\rangle_\alpha \label{eq:derivL}
\end{align}
Here :
\begin{equation}
{\bf F}^\alpha= \left(\widehat K +\widehat T^{-1}\left ({\bf x}^\alpha\right )\right)^{-1}  \frac{\partial \Lambda}{ \partial {\bf x}^{\alpha}_{\rm out}} \label{eq:y}
\end{equation}
and   $\Phi_{ij}^\alpha= \delta_{ij}F_i^{\alpha}$.
The set of vectors $\bf F^\alpha$  given by Eq. \ref{eq:y} can also be evaluated recursively, to avoid  inverting matrix $\widehat K +\widehat T^{-1}$ at each step, in combination with the recursive solution of Eqs. (\ref{eq:X}): 
\begin{align}
x_i^\alpha&\leftarrow\frac{c_i}{1+\sum_j K_{ij}x_j^\alpha} \label{eq:recurs1}\\
{\bf F}^\alpha&\leftarrow \widehat T ({\bf x}^\alpha) \left( \frac{\partial \Lambda}{ \partial {\bf x}^\alpha_{\it out}}-\widehat K {\bf F}^\alpha\right )\label{eq:recurs2}
\end{align}

We implemented  the gradient descent evolution in terms of concentration of the hidden layer $c_{\rm hid}$, and affinity parameters  $\tau_{ij}\equiv \ln \left(c_0K_{ij}\right)$, so that $c_0K_{ij}=\exp(\tau_{ij})$; these parameters are directly related to the corresponding binding free energies: $\tau_{ij}=-\Delta G_{ij}/RT$. The  two loss functions were  additionally amended by the regularization term:
\begin{equation}
    \widetilde{\Lambda}=\Lambda+\mu \sum_{i,j} \tau_{ij}
\end{equation}
The resulting evolution for the GD evolution of the hidden concentration and the affinity parameters could thus be obtained as
\begin{align}
c_{\rm hid} & \leftarrow  c_{\rm hid} - \epsilon_{c}\frac{\partial \widetilde{\Lambda} }{\partial c_{\rm hid}}= c_{\rm hid} -\epsilon_{c}\sum_{i\in {\rm hidden}} \left\langle\frac{F_i^\alpha}{x_i^\alpha}\right\rangle_\alpha\\
    \tau_{ij} & \leftarrow \tau_{ij} -\epsilon_\tau \frac{\partial \widetilde{\Lambda}}{\partial \tau_{ij}} = \tau_{ij} -\epsilon_\tau\left( K_{ij}\frac{\partial \Lambda }{\partial K_{ij}} + \mu\right)
\end{align}

Here $\partial \Lambda /\partial \widehat K$ and $F$ are determined  by Eqs.(\ref{eq:derivL}),(\ref{eq:recurs1}), (\ref{eq:recurs2}), which  in turn depend on computing the loss function  derivatives  $\frac{\partial \Lambda}{ \partial {\bf x}_{\it out}}$:
\begin{align}
    \frac{\partial \Lambda_{\it contrast}}{ \partial x^{\alpha}_i} =& \frac{\delta_{ij}}{N_Bc_i} \left(
    \frac{1-\phi^\beta_j}{\langle (1-\phi^\alpha_j) f^\alpha_j \rangle_\alpha} - 3 \frac{\phi^\beta_j}{f^\beta_j}
    \right),  \\
    & \nonumber  j=\mathrm{argmax} \{ \Lambda_{{\it contrast},j} \}\\
    \frac{\partial \Lambda_{\rm mse}}{\partial x^{\alpha}_i} =&\frac{2( f^\alpha_i - \phi^\alpha_i)} { N_Bc_i} 
\end{align}
In practice, instead of employing this iterative scheme, we use the adaptive learning rate algorithm from Ref. \cite{Adam}.

 \subsection{Transient behavior}
\begin{figure*}        \includegraphics[width=\linewidth]{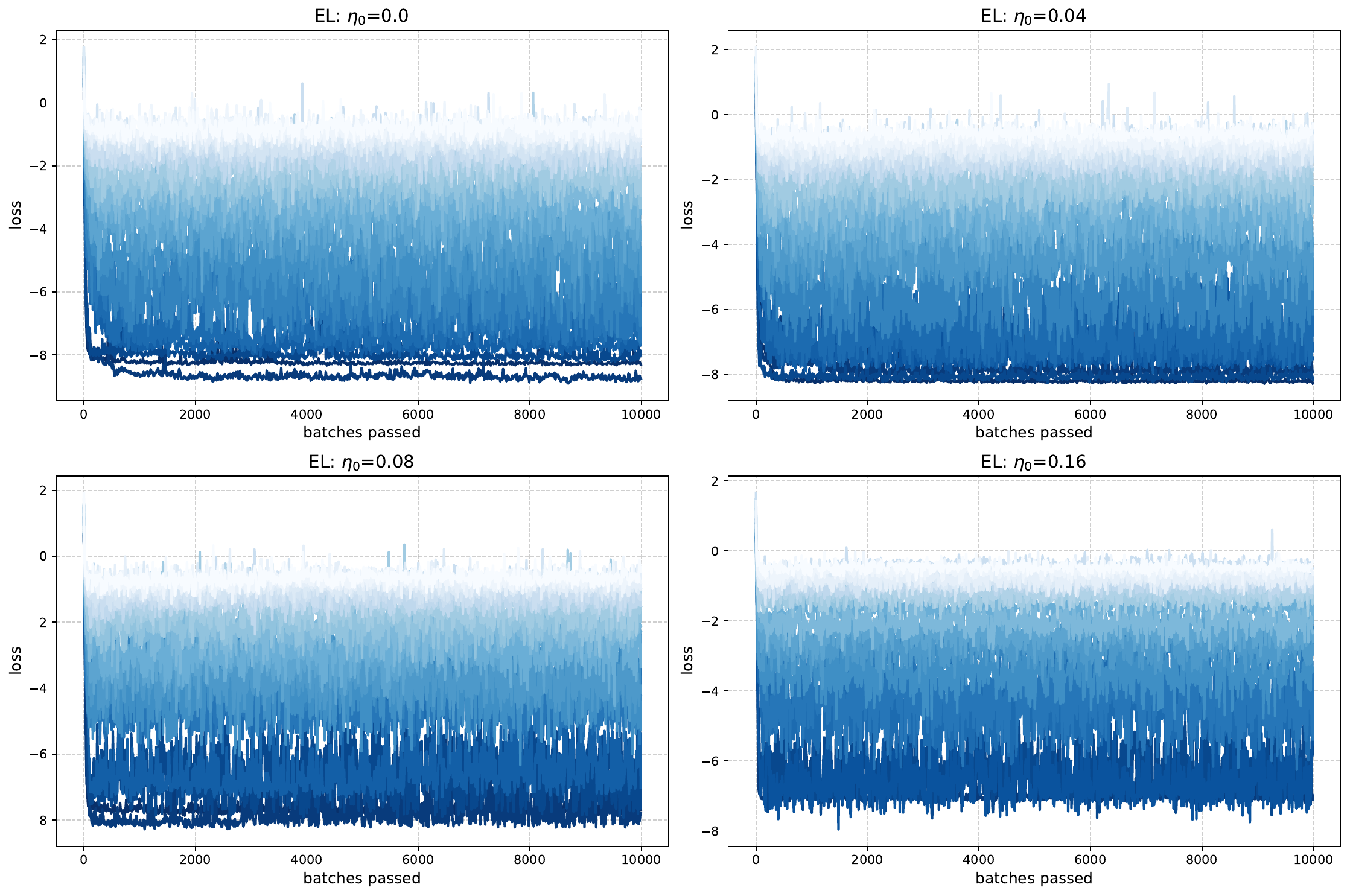}
        \caption{
        {\bf Evolution of the mean squared error (MSE) loss during training under the Evolutionary Learning (EL) algorithm.} Each panel corresponds to a different drift parameter $\eta_0$ (see Eq. 3 in the main text). Within each panel, curves represent 23 different noise-to-signal (N/S) ratios (10\%$\leq$N/S$\leq$120\%). Higher N/S ratios are indicated by lighter (more faded) colors.}
        \label{firstSI}
\end{figure*}

Figures \ref{firstSI}-\ref{lastSI}  illustrate how the loss functions and the performance metric evolve during EL and GD training. It is important to note that each step of EL is highly parallel, as $150$ different variants are exposed to $48$ different conditions. 
\begin{figure*}        \includegraphics[width=\linewidth]{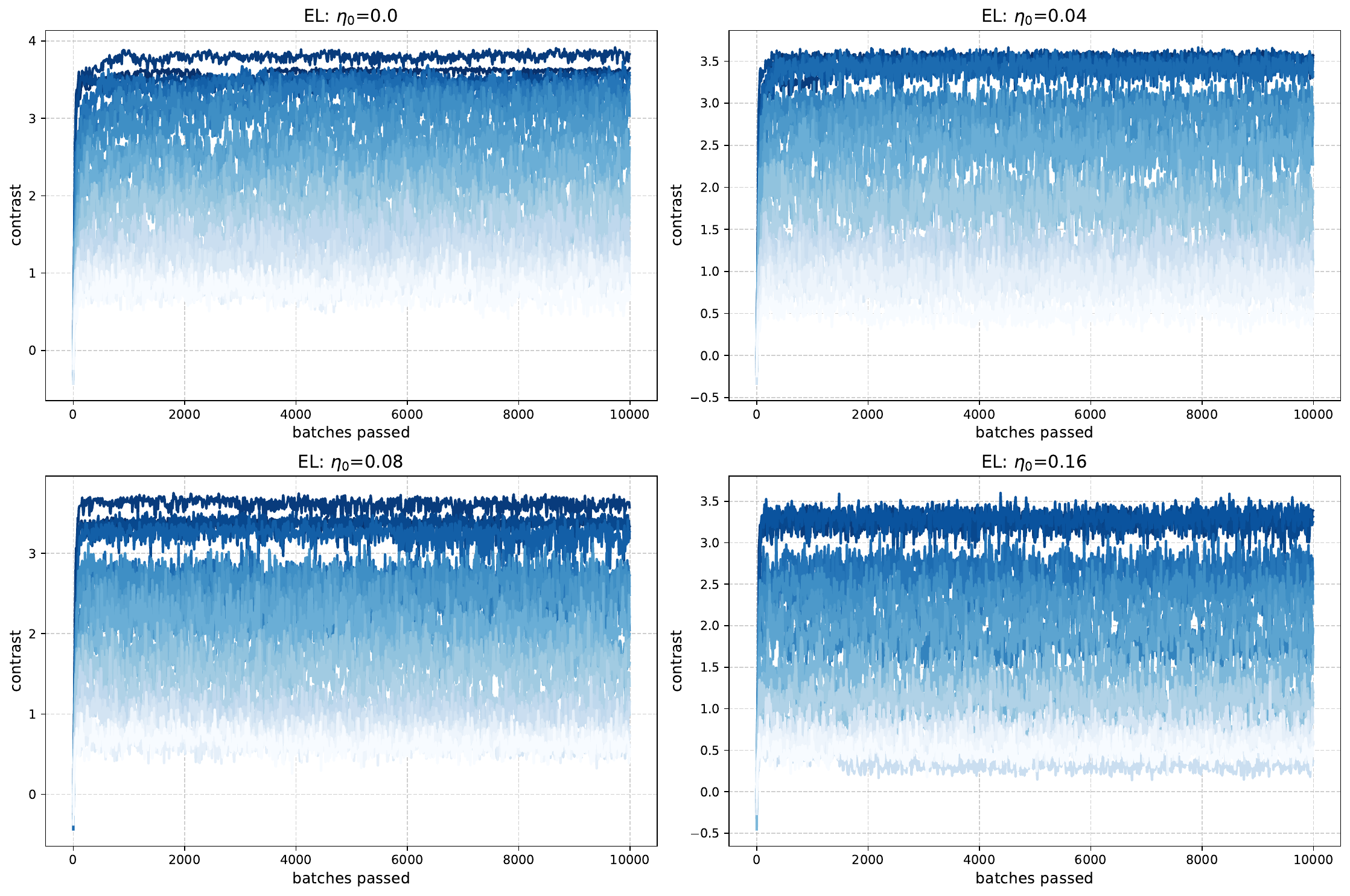}
        \caption{
        {\bf Evolution of the contrast loss (see Eq.\ 4 in the main text) during training under the Evolutionary Learning (EL) algorithm.} Each panel corresponds to a different drift parameter $\eta_0$ (see Eq. 3 in the main text). Within each panel, curves represent 23 different noise-to-signal (N/S) ratios (10\%$\leq$N/S$\leq$120\%). Higher N/S ratios are indicated by lighter (more faded) colors.}
\end{figure*}
\begin{figure*}        \includegraphics[width=\linewidth]{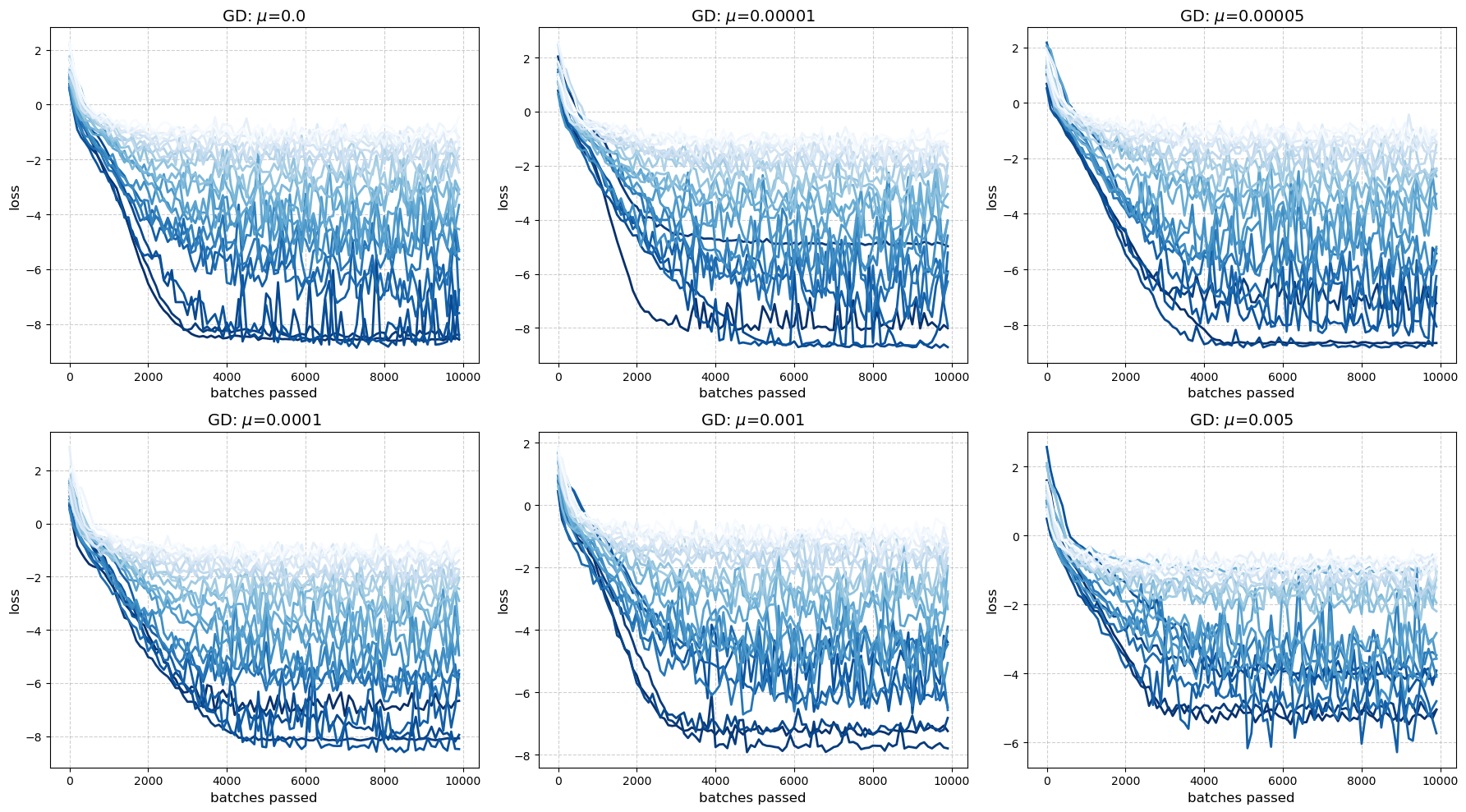}
        \caption{
        {\bf Evolution of the mean squared error (MSE) loss during training using the Gradient Descent  (GD) algorithm.} Each panel corresponds to a different regularization term $\mu$. Within each panel, curves represent 23 different noise-to-signal (N/S) ratios (10\%$\leq$N/S$\leq$120\%). Higher N/S ratios are indicated by lighter (more faded) colors.}
\end{figure*}
\begin{figure*}
\includegraphics[width=\linewidth]{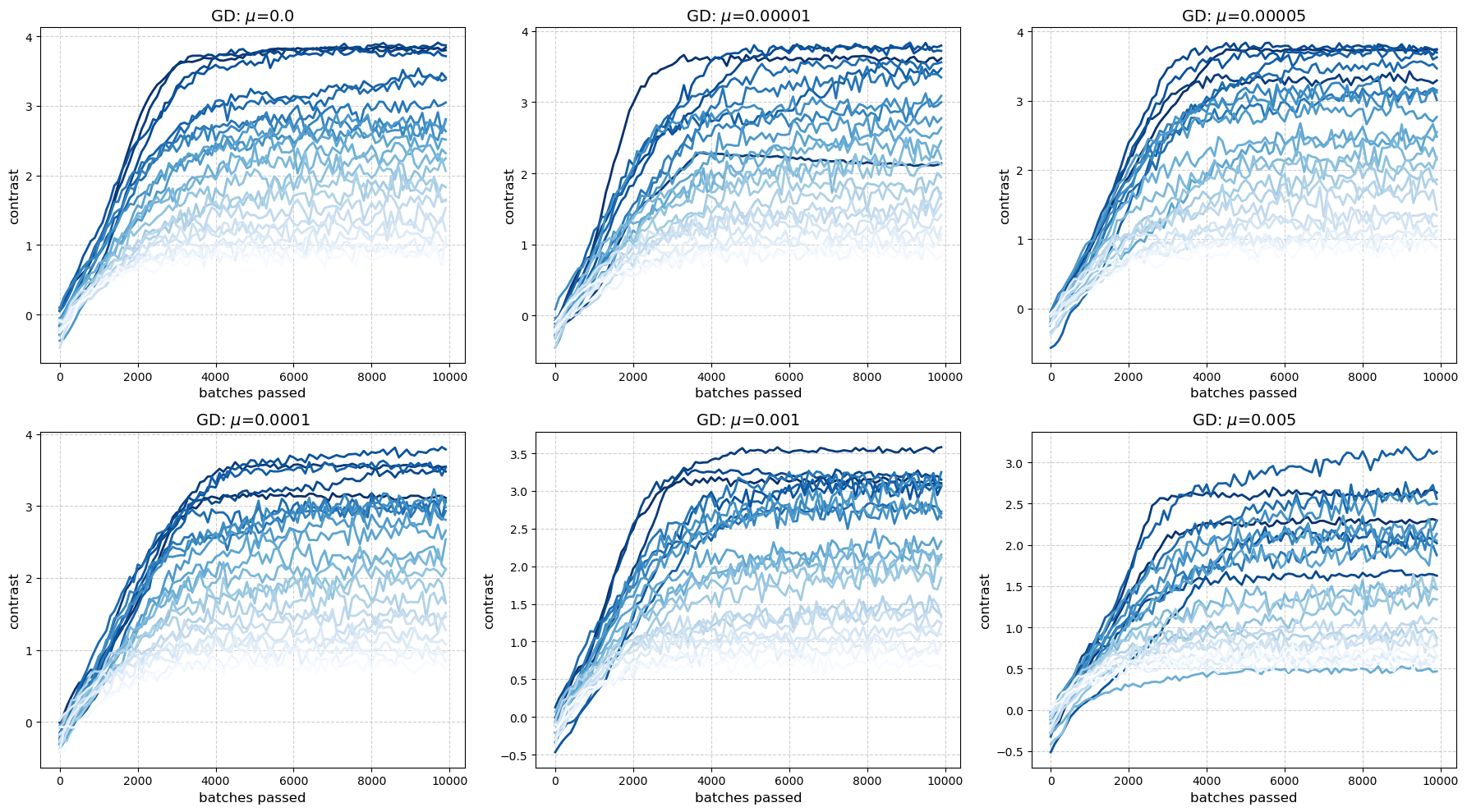}
        \caption{
        {\bf Evolution of the contrast loss (see Eq.\ 4 in the main text) during training using the Gradient Descent  (GD) algorithm.} Each panel corresponds to a different regularization term $\mu$. Within each panel, curves represent 23 different noise-to-signal (N/S) ratios (10\%$\leq$N/S$\leq$120\%). Higher N/S ratios are indicated by lighter (more faded) colors.}
        \label{lastSI}
\end{figure*}

\end{document}